\documentclass{article}

\usepackage{PRIMEarxiv}

\usepackage[utf8]{inputenc} 
\usepackage[T1]{fontenc}    
\usepackage[hidelinks]{hyperref}       
\usepackage{url}            
\usepackage{booktabs}       
\usepackage{amsfonts}       
\usepackage{nicefrac}       
\usepackage{microtype}      
\usepackage{lipsum}
\usepackage{fancyhdr}       
\usepackage{graphicx}       
\usepackage{amsmath}
\usepackage{multirow}
\usepackage{subcaption}
\graphicspath{{media/}}     

\title{Estimation of non-uniform motion blur using a patch-based regression convolutional neural network (CNN)
}

\author{
  Luis G.~Varela, Laura E. Boucheron, Steven Sandoval, David Voelz, and Abu Bucker Siddik \\
  Klipsch School of Electrical and Computer Engineering \\
  New Mexico State University \\
  Las Cruces, NM 88003\\
  \texttt{\{varelal, lboucher\}@nmsu.edu} \\
}

\begin{document}
\maketitle\let\thefootnote\relax\footnotetext{The official version of this paper can be accessed via \url{https://doi.org/10.1364/AO.521076}.\\ Copyright 2024 Optica Publishing Group. One print or electronic copy may be made for personal use only. Systematic reproduction and distribution, duplication of any material in this paper for a fee or for commercial purposes, or modification of the content of the paper are prohibited.}

\begin{abstract}
The non-uniform blur of atmospheric turbulence can be modeled as a superposition of linear motion blur kernels at a patch level. We propose a regression convolutional neural network (CNN) to predict angle and length of a linear motion blur kernel for varying sized patches. We analyze the robustness of the network for different patch sizes and the performance of the network in regions where the characteristics of the blur are transitioning. Alternating patch sizes per epoch in training, we find coefficient of determination scores across a range of patch sizes of $R^2>0.78$ for length and $R^2>0.94$ for angle prediction.  We find that blur predictions in regions overlapping two blur characteristics transition between the two characteristics as overlap changes.  These results validate the use of such a network for prediction of non-uniform blur characteristics at a patch level.
\end{abstract}


\section{Introduction}
The effects of atmospheric turbulence in an image are commonly modeled as a spatially-varying blur~\cite{atmospheric,thiebaut2016spatially} and spatially-varying blurred images can be modeled as a superposition of locally linear motion blurs~\cite{bahat,Sun_2015_CVPR,Kim_2014_CVPR}. Linear motion blur, i.e., motion blur that can be described by motion in a line, is commonly used as a model for camera shake, camera platform movement, and moving objects in images~\cite{fergus,hirsch_2011_ieee,FMD}.  In this work, we parametrize linear motion blur by length and angle and estimate those parameters in image patches using a regression convolutional neural network (CNN).  In previous work it was demonstrated that a regression CNN can accurately predict length and angle of motion blur in uniformly blurred images and that those predictions can be used in non-blind image deblurring methods to recover the latent sharp image~\cite{varela2023_uniform}.  The work in~\cite{varela2023_uniform} was extended in~\cite{varela2023_patch} to a patch-based approach to estimate non-uniform blur.  This paper extends the work in~\cite{varela2023_patch} with a comprehensive study of the effect of patch size on the accuracy of parameter estimation and estimation accuracy for patches that overlap different motion blur characteristics.

There are two main contributions of this work. First, a single network is trained to be robust to a range of patch sizes through a training approach that alternates patch size per epoch and by removing two convolution layers from the network in~\cite{varela2023_patch} to allow for stability in prediction for a range of patch sizes. This results in a network that can be used for prediction of non-uniform blur for a wide range of patch sizes, depending on the expected spatial scale of the blur and allowing for a tradeoff between prediction performance and localization of blur characteristics, and that alleviates the need to train multiple networks for multiple patch sizes which can be computationally intensive.  This is in contrast to most patch-based methods which are designed to operate on a fixed-size patch, usually approximately $30\time30$ pixels, e.g.,~\cite{Sun_2015_CVPR, bahat, yan_2016_ieee}.  Second, we study blur characteristic predictions in patches that overlap regions of two blur characteristics which, to the best of our knowledge, has not been studied before.  This provides important insights for the prediction of blur characteristics in non-uniformly blurred images for methods that operate on image patches.  Image patches can be expected to span a range of blur characteristics and blur transitions in non-uniformly blurred images and the specific performance of blur prediction networks in such scenarios is important to understand.

The remainder of this paper is organized as follows.  Section~\ref{sec:background} presents an overview of models for uniform and non-uniform image blurring and a discussion of related work for image blur estimation and removal.  Section~\ref{sec:patch_CNN} presents the patch-based regression CNN architecture, details of training, and an overview of the dataset used for training and testing.  Section~\ref{sec:results} discusses results, including the effects of patch size on length and angle estimation and effects of overlapping blur characteristics within patches.  Section~\ref{sec:conclusion} provides conclusions and a brief discussion of future work. 

\section{Background}
\label{sec:background}
\subsection{Image Blur}
A uniformly blurred image can be modeled as 
\begin{equation}
    \label{eq:uniform}
    I_{\text{b}} = K * I_{\text{s}} + \eta,
\end{equation}
where $I_{\text{b}}$ is the blurred image, $K$ is the blur kernel, $I_{\text{s}}$ is the sharp latent image, $\eta$ is additive noise, and $*$ is the convolution operator (or correlation operator as is commonly assumed in image filtering). This formulation assumes a uniform blur because the same kernel $K$ is applied across the entire image.  A non-uniform blur can be represented as~\cite{gong2017motion}
\begin{equation}
    I_{\text{b}}[m,n] = \sum_{s}\sum_{t} K_{m,n}[s,t]I_{\text{s}}[m+s,n+t] + \eta,
    \label{eq:nonuniform}
\end{equation}
where $K_{m,n}$ is a blur kernel centered at pixel $I_{\text{s}}[m,n]$, the summation over $s$ and $t$ are defined over the extent of the kernel, and the indices for sharp image $I_{\text{s}}$ implement a correlation (rather than a convolution).  Thus, in the most general case for Eq.~(\ref{eq:nonuniform}), each pixel $I_{\text{s}}$ has a different blur kernel operating on its neighborhood, defined by the extent of the blur kernel.  Note that if $K_{m,n}=K$, i.e., the same blur kernel is applied to all pixels, then the non-uniform blur in Eq.~(\ref{eq:nonuniform}) reduces to the uniform blur of Eq.~(\ref{eq:uniform}).

\subsection{Estimating Image Blur Parameters and Deblurring}
Many approaches to deblurring and blur kernel estimation prior to deep learning used a maximum \textit{a posteriori} (MAP) framework with implicit regularization using image priors~\cite{fergus,FMD,Radon_T,MONEY2008302,jia_2007_ieee} or explicit edge prediction~\cite{Xu_2013_CVPR,BD_Krishnan}.  The MAP approach is an iterative optimization with alternate iterations predicting the blur kernel and the latent sharp image.  Levin et al.~\cite{Levin_2011_ieee,Levin_2009_ieee}, however, posit that such an approach tends to favor a solution in which the kernel is an impulse and the ``deblurred'' image is the blurry image instead of converging to the true blur kernel and latent sharp image.  As such, Levin et al.~\cite{Levin_2011_ieee,Levin_2009_ieee} advocate for approaches that only estimate the blur kernel (they use an expectation-maximization (EM) framework with either a Gaussian or sparse prior) and use of non-blind deblurring methods to estimate the latent sharp image.  Having demonstrated the utility of using estimated blur kernels in non-blind deblurring in our previous work on uniformly blurred images~\cite{varela2023_uniform}, here we focus on estimation of the blur kernel in non-uniformly blurred images.

Recent methods have considered deep learning approaches to estimate the blur kernel or blur field~\cite{yan_2016_ieee,Sun_2015_CVPR,Xu_2018_ieee,gong2017motion}, estimate the latent sharp image~\cite{zhang_2018_ieee}, or estimate both the blur kernel or blur field and the latent sharp image~\cite{Li_2018_CVPR} or compensated image~\cite{kageyama2022,kageyama2024}.  These methods use CNNs~\cite{Xu_2018_ieee,Sun_2015_CVPR,Li_2018_CVPR,yan_2016_ieee,kageyama2022,kageyama2024}, fully convolutional networks (FCNs)~\cite{gong2017motion,kageyama2022}, or generative adversarial networks (GANs)~\cite{zhang_2018_ieee}. CNNs have been used to enhance image edges as input to a classical MAP approach for uniformly-blurred images~\cite{Xu_2018_ieee}, to classify whether an image is blurry or sharp for use as regularization in a MAP framework~\cite{Li_2018_CVPR}, to predict defocus and to deblur images~\cite{kageyama2022,kageyama2024}, to predict blur kernels via classification for uniformly-blurred image patches~\cite{Sun_2015_CVPR}, and to classify the blur type (including defocus blur) followed by a general regression network to estimate the blur parameters~\cite{yan_2016_ieee}.  In contrast, we use a regression CNN to directly estimate the blur kernel or blur field by estimating motion blur parameters in patches.  

In this work we extend the regression CNN of~\cite{varela2023_uniform,varela2023_patch} which can be trained as a regressor on image patches to estimate length and angle as a parametrization of linear motion blur kernels. The use of regression in this formulation allows us to explore a continuous range of length and angle parameters for linear motion blur. This work improves the results from~\cite{varela2023_patch} by introducing an alternate patch size training. Furthermore, we create non-uniformly blurred images by blurring two regions of the image with different characteristics and use the network to predict blur in patches that overlap the two blur characteristics.  This gives important insights into the behavior of the network when analyzing patches that include multiple blur characteristics.  It is important to note that the network is trained on patches with uniform blur and is not explicitly trained on patches containing multiple blur characteristics; it is thus critical to understand the performance of the network in such circumstances.

\section{Patch-Based Regression Convolutional Network}
\label{sec:patch_CNN}
\subsection{Motion Blur Parametrization}
\label{sec:parametrization}
We parametrize linear motion blur kernels by the magnitude of the blur length, $r$, and the orientation angle of the blur, $\phi$.  In previous work~\cite{varela2023_uniform}, exploration of a suite of length and angle pairs $(r,\phi)$, $r\in[1,100]$, $\phi\in(-90,90]$ demonstrated that not all $(r,\phi)$ combinations result in a unique blur kernel with shorter length kernels tending to have fewer unique angles. The COCO dataset~\cite{lin2014microsoft} was used to create a blurred dataset with the collection of unique linear motion blur kernels with $(r,\phi)=(1,0)$ used to designate a non-blurred image. A modified version of VGG16~\cite{vgg16} was trained, where the last fully connected layer was modified to have two outputs (for length and angle) and to predict values in the range $[0,1]$ using a sigmoid activation for normalized values of $r$ and $\phi$.  Here, we consider a modified range of length $r$ to be commensurate with the maximum patch size used in training, i.e., $r\in[1,\min(N_\textrm{max},100)]$, where $N_\textrm{max}$ is the maximum patch size used in training and 100 is the maximum blur length considered in~\cite{varela2023_uniform}.  The need for this constraint is further discussed in Section~\ref{sec:training}.

\subsection{Network Architecture}
As in~\cite{varela2023_patch}, in order to input dimensions other than a fixed $224\times224$ pixels, we replace the flattening layer in the VGG16 network with a global average pooling (GAP) layer.  To mitigate issues associated with the fewer outputs of a GAP layer compared to a fully connected layer, the number of filters in later convolutional layers is increased and the number of nodes in the fully connected layers is decreased.  This results in the network architecture shown in Table~\ref{table: VGG16}, the same architecture as used in~\cite{varela2023_patch}.  While this network can accept an arbitrary size input patch of $N\times N$ pixels, $N$ is lower bounded by 32.  This is due to Block 5 having an output size of $N/32$ since each block in the model is followed by a max pooling layer with stride $(2,2)$. Here, we remove Block 1 to lower the bound of minimum patch size to 16. The same simulation from~\cite{varela2023_patch} trained on patch size 112 was run with the removal of Block 1 and we noted no substantial changes in results (see Table~\ref{tab:r2_length} for results).    

\begin{table}[t]
\centering
\caption{Modified VGG16 model used for patch-based estimation.}  
\begin{tabular}{lll}
\toprule
\textbf{Layer}   & \textbf{Description}   & \textbf{Output Dimensions}\\
\midrule
Input   & Input         & $N\times N\times3$   \\\hline
Block 1 & $2\times$Conv & $(N/2)\times(N/2)\times64$   \\\hline
Block 2 & $2\times$Conv & $(N/4)\times(N/4)\times128$  \\\hline
\multirow{2}{*}{Block 3} & $1\times$Conv  & $(N/8)\times(N/8)\times256$  \\
                         & $2\times$Conv  & $(N/8)\times(N/8)\times512$  \\\hline
Block 4 & $3\times$Conv  & $(N/16)\times(N/16)\times1024$  \\\hline
Block 5 & $3\times$Conv  & $(N/32)\times(N/32)\times2048$  \\\hline
Block 6 & Global Avg Pool & $1\times2048$  \\\hline
FC1     & Fully Connected & 2048    \\\hline
FC2     & Fully Connected & 2048    \\\hline
FC3     & Output          & 2  \\
\bottomrule
\multicolumn{3}{p{3.15in}}{\footnotesize{Notes: Each block is followed by a max pooling layer with stride $(2,2)$.  All convolutional and fully connected layers have a ReLU activation except the final fully connected layer FC3 which has a sigmoid activation.  Output dimensions for convolution layers are described as length $\times$ width $\times$ channels and all remaining layers are one-dimensional.}}
\end{tabular}
\label{table: VGG16}
\end{table}

\subsection{Training}
\label{sec:training}
The training of the network is similar to~\cite{varela2023_patch} which trains the network from scratch with TensorFlow's default layer initializations, uses the Adam optimizer with a learning rate of $0.1$ and epsilon of $0.1$, and uses the mean squared error (MSE) as the loss function. The training is run for $100$ epochs with early termination if no improvement in the loss function is seen within $15$ epochs. A few modifications adjust the training for different patch sizes.  First, the normalization of $r$ is modified to be consistent with the maximum trained blur length $N_\textrm{max}$ to utilize the full range of the sigmoid activation.  Second, we implement training wherein we change the input patch size per epoch.  As described in Section~\ref{sec:results}, this allows the network to specialize on specific patch sizes within an epoch and improve performance across a range of patch sizes.  Thus, changing the input patch size per epoch results in a model that is more robust in predicting blur characteristics across a wider range of patch sizes.  We use the blurred COCO dataset described in~\cite{varela2023_uniform,Regression_Blur_github} to train and test the results of the architecture; subsets of the dataset are used as described below.

Training the network with different patch sizes per epoch requires modification of the data generator used in training.  Since the blurred COCO dataset described in~\cite{varela2023_uniform,Regression_Blur_github} contains images blurred with kernels of length $r\in[1,00]$, images blurred with blur parameters that will not fit within the patch size $N$ must be removed in training.  For example a $32\times32$ pixel patch can accommodate a length-32 horizontal blur, but could accommodate a length-$32\sqrt{2}$ blur at an angle of 45 degrees.  As in~\cite{varela2023_patch}, we keep blurred images with $(r,\phi)$ such that  
\begin{equation}
    r \le \begin{cases}
        \sqrt{N^2(1 + \tan^2(|\phi|)},  |\phi| \le 45^{\circ}
        \\
        \sqrt{N^2(1 + \cot^2(|\phi|)},  |\phi| > 45^{\circ}
        \end{cases},
        \label{eq:max_length}
\end{equation}
where $N$ is the patch size. In short, this equation keeps blurred images from the training set only if the lengths are short enough for the patch size to accommodate by recognizing whether the horizontal or vertical axis uses the full $N$ pixels and using a simple application of the Pythagorean theorem.  Here, blurred images from the training dataset are dynamically removed in each epoch since the patch size $N$ may change per epoch. Implementation of Eq.~(\ref{eq:max_length}) is added to the data generator in training, ensuring that only those data that satisfy Eq.~(\ref{eq:max_length}) are seen in training. The data generator creates batches of samples consisting of matched pairs of image patches and normalized labels by randomly selecting from the entire dataset.  For each sample, the labels are denormalized (restored to their original range) and if the denormalized length $r$ satisfies Eq.~(\ref{eq:max_length}), the sample is appended to the batch.  This implementation of appending samples which satisfy Eq.~(\ref{eq:max_length}) is important, as removal of samples from a standard data generator batch can result in empty batches (which we found resulted in the loss function returning not-a-number (NaN) and training not converging).  

We desire a single network that is robust to different patch sizes for maximum flexibility in prediction of motion blur.  For example, this single network could predict blur characteristics at different granularities or be used for prediction of blur in different levels of turbulence which are expected to manifest on different length scales.  Furthermore, we prefer that the network be most accurate for prediction of blur characteristics at smaller patch sizes, as we expect that will prove most accurate in the restoration of a non-uniformly blurred image.  Some limited generalization to other patch sizes was seen in~\cite{varela2023_patch}, also summarized in Table~\ref{tab:r2_length}, where both length and angle predictions were accurate for larger patches and accuracy suffered for patches smaller than half the trained size.  To this end we modified the data generator to accept an array of patch sizes $P$ to specify the patch size per epoch.  The array of patch sizes is indexed circularly, looping back to the first element after the last element.  For example, a patch size array of $P=[32, 64, 112, 224]$ would result in epochs being trained until convergence (change in the loss function is $<10^{-12}$ for $15$ epochs) on patch sizes of $32, 64, 112, 224, 32\ldots$.   

\subsection{Data}
\label{sec:data}
\begin{figure}[t!]
    \centering
    \subcaptionbox{}{\includegraphics[width=0.3\textwidth]{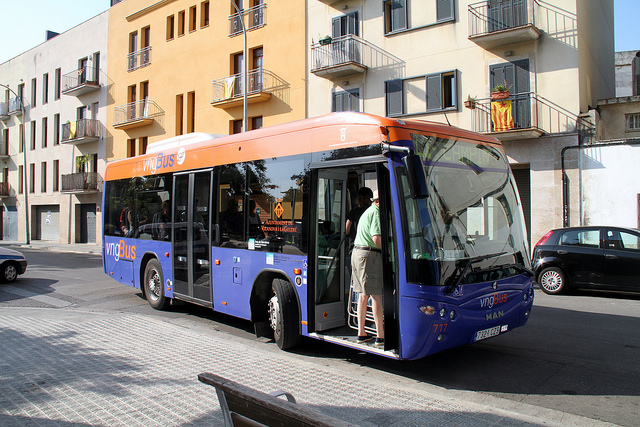}}~~~~~
    \subcaptionbox{}{\includegraphics[width=0.3\textwidth]{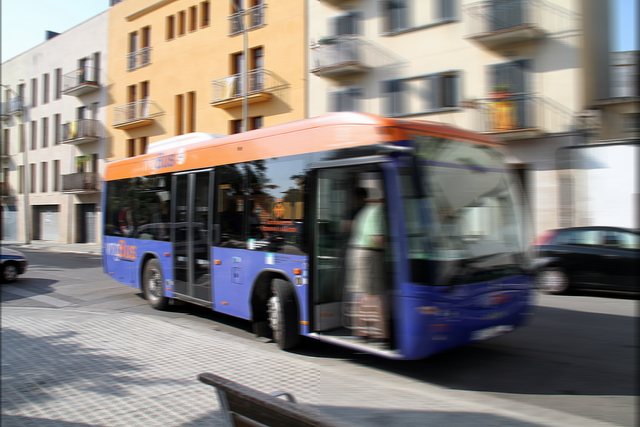}}~~~~~
    \subcaptionbox{}{\includegraphics[width=0.3\textwidth]{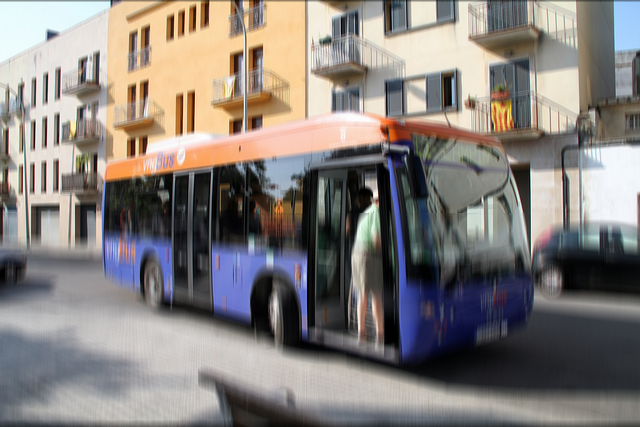}}\\
    \subcaptionbox{}{\includegraphics[width=0.3\textwidth]{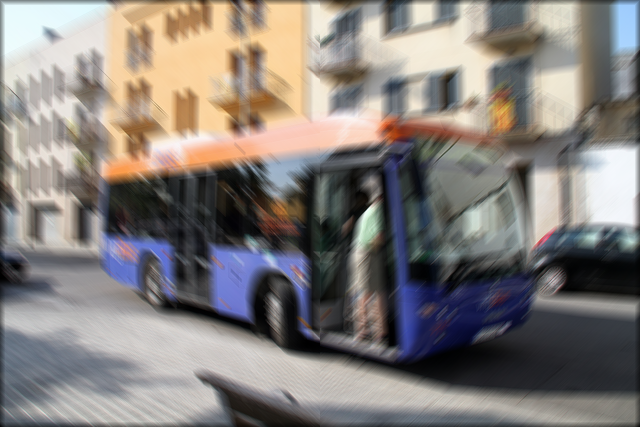}}~~~~~
    \subcaptionbox{}{\includegraphics[width=0.3\textwidth]{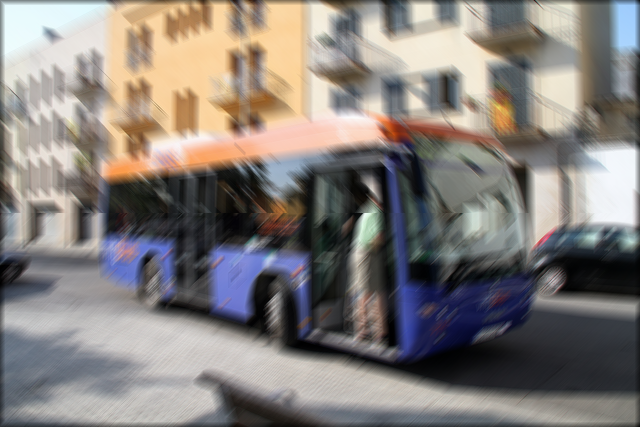}}~~~~~
    \caption{Examples of non-uniformly blurred images.  (a) Original unblurred image.  (b) \textit{Length Horizontal}: $\phi=0$ and $r=(5,15)$ in the (left, right). (c) \textit{Length Vertical}: $\phi=90$ and $r=(5,15)$ in the (top, bottom). (d) \textit{Angle Horizontal}: $r=15$ and $\phi=(-45,45)$ in the (left, right).  (e) \textit{Angle Vertical}: $r=15$ and $\phi=(-45,45)$ in the (top, bottom).  }
    \label{fig:nonuniform}
\end{figure}
The blurred COCO dataset previously described in~\cite{varela2023_uniform,Regression_Blur_github} is used to test robustness of the network to patch size (Section~\ref{sec:patch_size}).  To test the network in cases where the patch overlaps regions of different blur characteristics (Section~\ref{sec:overlap}), a new non-uniformly blurred dataset is created, again using the COCO dataset~\cite{lin2014microsoft}.  Here, we consider four cases of nonuniform blur, also illustrated in Fig.~\ref{fig:nonuniform}.  Fig.~\ref{fig:nonuniform}(b) illustrates the case where length is varied horizontally, i.e., $\phi=0$, with the length changing from $r=5$ in the left half of the image to $r=15$ in the right; we subsequently refer to images blurred with this blur field as \textit{Length Horizontal}.  Fig.~\ref{fig:nonuniform}(c) illustrates the case where length is varied vertically, i.e., $\phi=90$, with the length changing from $r=5$ in the top half to $r=15$ in the bottom (\textit{Length Vertical}).  Fig.~\ref{fig:nonuniform}(d) illustrates the case where angle is varied horizontally, with length $r=15$ and the angle changing from $\phi=-45$ in the left half to $\phi=45$ in the right (\textit{Angle Horizontal}).  Fig.~\ref{fig:nonuniform}(e) illustrates the case where angle is varied vertically, with length $r=15$ and the angle changing from $\phi=-45$ in the top half to $\phi=45$ in the bottom (\textit{Angle Vertical}).  These non-uniformly blurred images allow us to study the effects of prediction of blur characteristics as image patches overlap with different blur characteristics (Section~\ref{sec:overlap}), a critical consideration since the network is trained only on uniformly blurred images.  

\section{Results and Discussion}
\label{sec:results}
\subsection{Performance Metric}
As in~\cite{varela2023_uniform,varela2023_patch}, we use the coefficient of determination $R^2$ score~\cite{r2} to measure the performance of the model in predicting length and angle of blur from blurry images. A more accurate model will have an $R^2$ closer to $1$, a na\"{i}ve model that predicts the data average will have an $R^2$ of $0$, and models with predictions worse than the na\"{i}ve model will have an $R^2$ score $<0$.

\subsection{Effects of Patch Size on Length and Angle Estimation}
\label{sec:patch_size}
Table~\ref{tab:r2_length} shows the results of the training simulations from~\cite{varela2023_uniform,varela2023_patch} as a baseline and new simulations with alternating patch size training as described above. First, we test the effects of removing Block 1 from VGG16 which is needed in order to train and test on patch sizes less than 32 pixels.  We train VGG16 without Block 1 on a patch size of 112, allowing us to directly compare the results to those of full VGG16 trained on patch size 112 as reported in~\cite{varela2023_patch}.  We note comparable performance for length prediction and slightly improved performance on angle prediction when tested on patch sizes of 64, 112, and 224.  We see a downward trend in the $R^2$ score when testing on patch sizes smaller than the trained size of 112, similar to the results seen in~\cite{varela2023_patch}. Since the removal of Block 1 has not detrimentally affected prediction on larger patch sizes and additionally  allows training and testing on patch sizes down to $16$, all new results in Table~\ref{tab:r2_length} are for an architecture without Block 1.  

\begin{table}[t!]
    \centering
    \caption{Coefficient of determination $R^2$ results for prediction of length $r$ and angle $\phi$ for different training configurations.}
    \small
    \begin{tabular}{lrrrrrrrr}    
      \multicolumn{6}{l}{Length $r$}\\
      \toprule
      & \multicolumn{8}{c}{\textbf{Test Patch Size}}\\ 
      \cmidrule(lr){2-9}
      \textbf{Training Config.} & \multicolumn{1}{c}{\textbf{16}} & \multicolumn{1}{c}{\textbf{29}} & \multicolumn{1}{c}{\textbf{30}} & \multicolumn{1}{c}{\textbf{31}} & \multicolumn{1}{c}{\textbf{32}} & \multicolumn{1}{c}{\textbf{64}} & \multicolumn{1}{c}{\textbf{112}} & \multicolumn{1}{c}{\textbf{224}} \\
      \midrule
      224~\cite{varela2023_uniform} & \multicolumn{1}{c}{-} & \multicolumn{1}{c}{-} & \multicolumn{1}{c}{-} & \multicolumn{1}{c}{-} & \multicolumn{1}{c}{-} & \multicolumn{1}{c}{-} & \multicolumn{1}{c}{-} & 0.987\\
      224~\cite{varela2023_patch} & \multicolumn{1}{c}{-} & \multicolumn{1}{c}{-} & \multicolumn{1}{c}{-} & \multicolumn{1}{c}{-} & -1.072 &  0.314 & 0.798 & 0.978\\
      112~\cite{varela2023_patch} & \multicolumn{1}{c}{-} & \multicolumn{1}{c}{-} & \multicolumn{1}{c}{-} & \multicolumn{1}{c}{-} & -0.427 &  0.748 & 0.946 & 0.986\\
      112 & -29.651 & \multicolumn{1}{c}{-} & \multicolumn{1}{c}{-} & \multicolumn{1}{c}{-} & -1.275 &  0.762 & 0.947 & 0.980\\
      31 & -1.020 & 0.563 & -0.802 & 0.821 & -0.360 &  0.399 & 0.548 & 0.610\\
      32, 64, 112, 224 &-1.145 & \multicolumn{1}{c}{-} & \multicolumn{1}{c}{-} & \multicolumn{1}{c}{-} & 0.594 & 0.865 & 0.958 & 0.985 \\
      30, 32, 32, 48, 64 & -0.657 & 0.734 & 0.765 & 0.752 & 0.815 & 0.358 & 0.426 & 0.447 \\
      29, 30, 31, 32, 33 & 0.129 & 0.786 & 0.787 & 0.788 & 0.837 & 0.847 & 0.831 & 0.810\\
      \bottomrule
      \\

      \multicolumn{6}{l}{Angle $\phi$}\\
      \toprule
      & \multicolumn{8}{c}{\textbf{Test Patch Size}}\\
      \cmidrule(lr){2-9}
      \textbf{Training Config.} & \multicolumn{1}{c}{\textbf{16}} & \multicolumn{1}{c}{\textbf{29}} & \multicolumn{1}{c}{\textbf{30}} & \multicolumn{1}{c}{\textbf{31}}& \multicolumn{1}{c}{\textbf{32}} & \multicolumn{1}{c}{\textbf{64}} & \multicolumn{1}{c}{\textbf{112}} & \multicolumn{1}{c}{\textbf{224}} \\
      \midrule
      224~\cite{varela2023_uniform} & \multicolumn{1}{c}{-} & \multicolumn{1}{c}{-} & \multicolumn{1}{c}{-} & \multicolumn{1}{c}{-}  & \multicolumn{1}{c}{-} & \multicolumn{1}{c}{-} & \multicolumn{1}{c}{-} & 0.994\\
      224~\cite{varela2023_patch} & \multicolumn{1}{c}{-} & \multicolumn{1}{c}{-} & \multicolumn{1}{c}{-} & \multicolumn{1}{c}{-} & 0.586 & 0.886 & 0.962 & 0.992\\
      112~\cite{varela2023_patch} & \multicolumn{1}{c}{-} & \multicolumn{1}{c}{-} & \multicolumn{1}{c}{-} & \multicolumn{1}{c}{-} & 0.751 & 0.748 & 0.946 & 0.986\\
      112 & 0.317 & \multicolumn{1}{c}{-} & \multicolumn{1}{c}{-} & \multicolumn{1}{c}{-} & 0.822 &  0.968 & 0.987 & 0.997\\
      31 & 0.301 & 0.917 & 0.853 & 0.954 & 0.524 &  0.690 & 0.719 & 0.715\\
      32, 64, 112, 224 & 0.588 & \multicolumn{1}{c}{-} & \multicolumn{1}{c}{-} & \multicolumn{1}{c}{-} & 0.955 & 0.993 & 0.997 & 0.999 \\
      30, 32, 32, 48, 64 & 0.706 & 0.941 & 0.948 & 0.950 & 0.964 & 0.983 & 0.988 & 0.991 \\
      29, 30, 31, 32, 33 & 0.731 & 0.942 & 0.942 & 0.942 & 0.962 & 0.980 & 0.977 & 0.972 \\
      \bottomrule
      \multicolumn{9}{p{4.5in}}
      {\footnotesize{Notes: All results are tested on the entire test subset (21,789 blurry images) of the blurred COCO dataset~\cite{varela2023_uniform,Regression_Blur_github}.  Training configurations are presented as the order in which patch sizes are presented per epoch (see Section~\ref{sec:training}).  A dash (-) indicates that the network was not tested on that patch size. Results from~\cite{varela2023_uniform,varela2023_patch} were trained and tested on a modified VGG16~\cite{vgg16} architecture (see Table~\ref{table: VGG16}) and new results use the modified VGG16 architecture with Block 1 removed.}}
    \end{tabular}
    \label{tab:r2_length}
\end{table}

Second, we introduce an alternating patch size training configuration as discussed in Section~\ref{sec:training} with the hypothesis that this will allow the network to focus on smaller patch sizes and result in a more uniform $R^2$ score when tested across different patch sizes. From the results in Table~\ref{tab:r2_length} we see that a network trained on a single patch size has good performance for that single patch size, but degraded performance for other patch sizes.  Since blur characteristics may manifest on a variety of length scales, the optimal patch size for blur prediction is not clear, and larger patch sizes will reduce the ability to localize blur characteristics, we desire a network that is robust to prediction at different patch sizes.  Additionally, it is important to remember that the network training is computationally intensive, requiring some 12--18 hours of training time on an NVIDIA RTX-3090.  A network that is robust to patch sizes will provide flexibility in application to images with different blur characteristics and computational simplicity.  As an additional baseline, we train on a fixed patch size of 31.  As seen in Table~\ref{tab:r2_length}, the network again struggles in predicting blur characteristics for patch sizes other than 31.

With a training configuration of $P=[32, 64, 112, 224]$, we see improved performance on smaller patch sizes compared to a network trained on one patch size of 112. The downward trend in $R^2$ score is still noticeable as patch size decreases, but the overall range from the best and worst $R^2$ score has diminished compared to training with a single patch size. Noting the lower performance on smaller patch sizes, we trained a model with a training configuration of $P=[30,32,32,48,64]$ with the hypothesis that the network would be forced to better predict smaller patch sizes.  Results demonstrate improved performance on smaller patch sizes (29, 30, 31, 32), but reduced performance on larger patch size (64, 112, 224).  Finally, we trained a network with a configuration of $P=[29,30,31,32,33]$ to specialize on patch sizes around 31 in order to test the effect of predicting blur characteristics in overlapping blur patches (Section~\ref{sec:overlap}).  We note a much more uniform performance for both length and angle prediction across a wide range of patch sizes, even up to 224.  Interestingly, it appears that important characteristics of blur can be learned within a relatively small image patch and generalized to larger image patches. Although the network is trained to simultaneously predict length and angle there seems to be a unique trend where length is more susceptible to the patch size compared to angle. In the next section we will analyze the behavior of prediction for each parameter as we transition into a non-uniform blur prediction region. It is this final configuration of $P=[29,30,31,32,33]$ that we use as the trained network for subsequent results.

\subsection{Analyzing Prediction of Blur Characteristics in Overlapping Patches}
\label{sec:overlap}
We study the effect of the patch overlapping different blur characteristics with the non-uniform blur patterns discussed in Section~\ref{sec:data}.  By definition, nonuniformly blurred images have blur characteristics that change spatially which leads to the question of how the network will predict blur in patches that contain contributions from different blur characteristics. Studying the behavior of the network around the boundary of two different blur characteristics is important since the network is being presented with image patches containing blurs unseen in the original training set. We randomly select 10 images from the COCO dataset and apply the four non-uniform blur patterns to each image.  Those 10 blurred images are input to the trained network (configuration $P=[29,30,31,32,33]$ as summarized in Table~\ref{tab:r2_length}) and length $r$ and angle $\phi$ are predicted for patches with size $31$ and advance of 1 pixel in a direction perpendicular to the blur discontinuity.  As an example, for the \textit{Length Horizontal} and \textit{Angle Horizontal} blurs, the network begins predicting characteristics for all patches at the left of the image, advancing the patch position to the right in steps of one pixel, resulting in varying degrees of overlap across the different blur characteristics.  Similarly, for \textit{Length Vertical} and \textit{Angle Vertical}, the network predicts blur characteristics for patches at the top of the image, advancing down in steps of one pixel.

\begin{figure}[t!]
        \centering
        \subcaptionbox{}{\includegraphics[width=\textwidth]{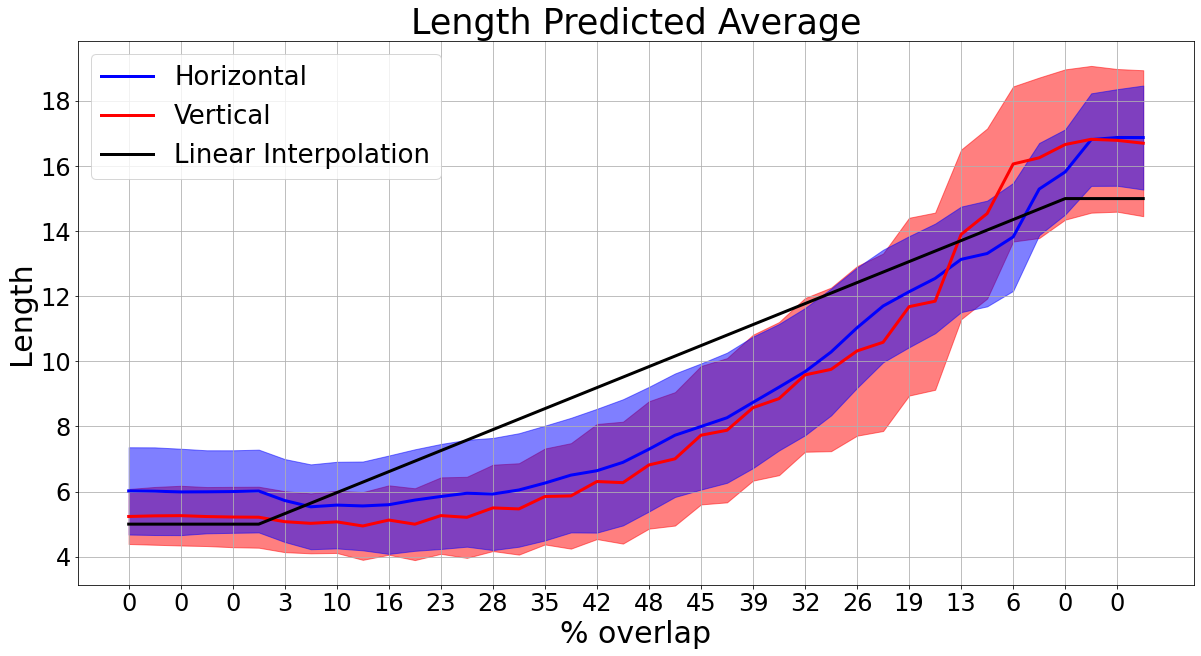}}\\ 
        \subcaptionbox{}{\includegraphics[width=\textwidth]{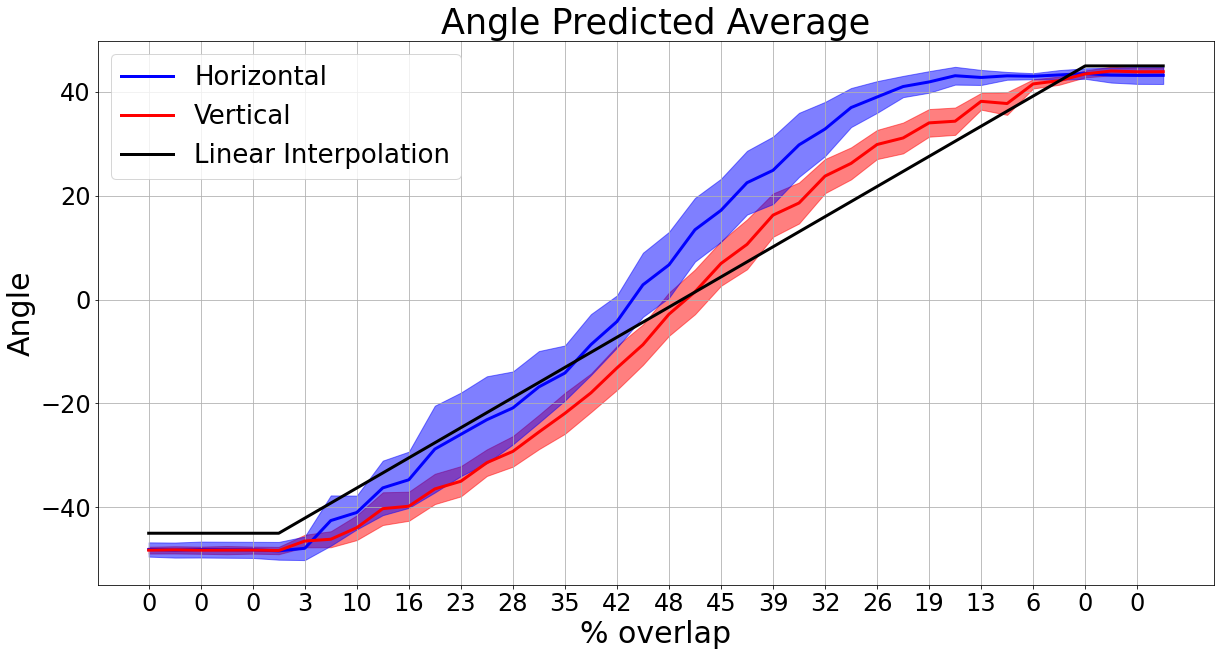}}\\
        \caption{Predicted blur characteristics versus percentage overlap for (a) changing length and (b) changing angle. The solid black line is the linear interpolation between the two blur characteristics as a point of reference.  The solid red and blue lines indicate average predictions across all $31\times31$ patches in the 10 non-uniformly blurred COCO images and the filled region is the corresponding standard deviation.  The x-axis is labeled from $0\%$ to $48\%$ overlap with the left side denoting leftward or upward regions and the right side denoting rightward and downward regions.}
        \label{fig:avg_plots}
\end{figure}

When cropping a patch from an image with overlap between two blur characteristics, we intuitively expect the prediction to be between the two blur characteristics.  Since the network has not seen such patches, however, the specific behavior should be carefully considered. Results for predicted characteristics for all $31\times31$ patches in the 10 non-uniformly blurred COCO images are shown in Fig.~\ref{fig:avg_plots}(a) for \textit{Length Horizontal} and \textit{Length Vertical} and Fig.~\ref{fig:avg_plots}(b) for \textit{Angle Horizontal} and \textit{Angle Vertical}. Additionally, in Fig.~\ref{fig:avg_plots}, we plot the linear interpolation between the two blur characteristics as a point of reference.  The results for change in angle [Fig.~\ref{fig:avg_plots}(b)] show average predictions with small variation and a symmetric transition relative to the overlap.  The results for change in length [Fig.~\ref{fig:avg_plots}(a)] show average predictions with larger variation and less symmetry relative to the overlap, but display a trend from shorter to longer lengths from left to right (the horizontal case) or top to bottom (the vertical case). We present here results for prediction of blur characteristics using a $31\times31$ patch for computational considerations (each of the 40 non-uniformly blurred images requires $\sim275$k predictions) and to avoid confounding effects from patch size and blur characteristics, but note that the same conclusions are expected to hold for other patch sizes.

\begin{figure}[t!]
        \centering
        \subcaptionbox{}{{\includegraphics[width=0.3\textwidth]{Img_Example/CLH_000000000139.png}}~~~~~
        {\includegraphics[width=0.3\textwidth]{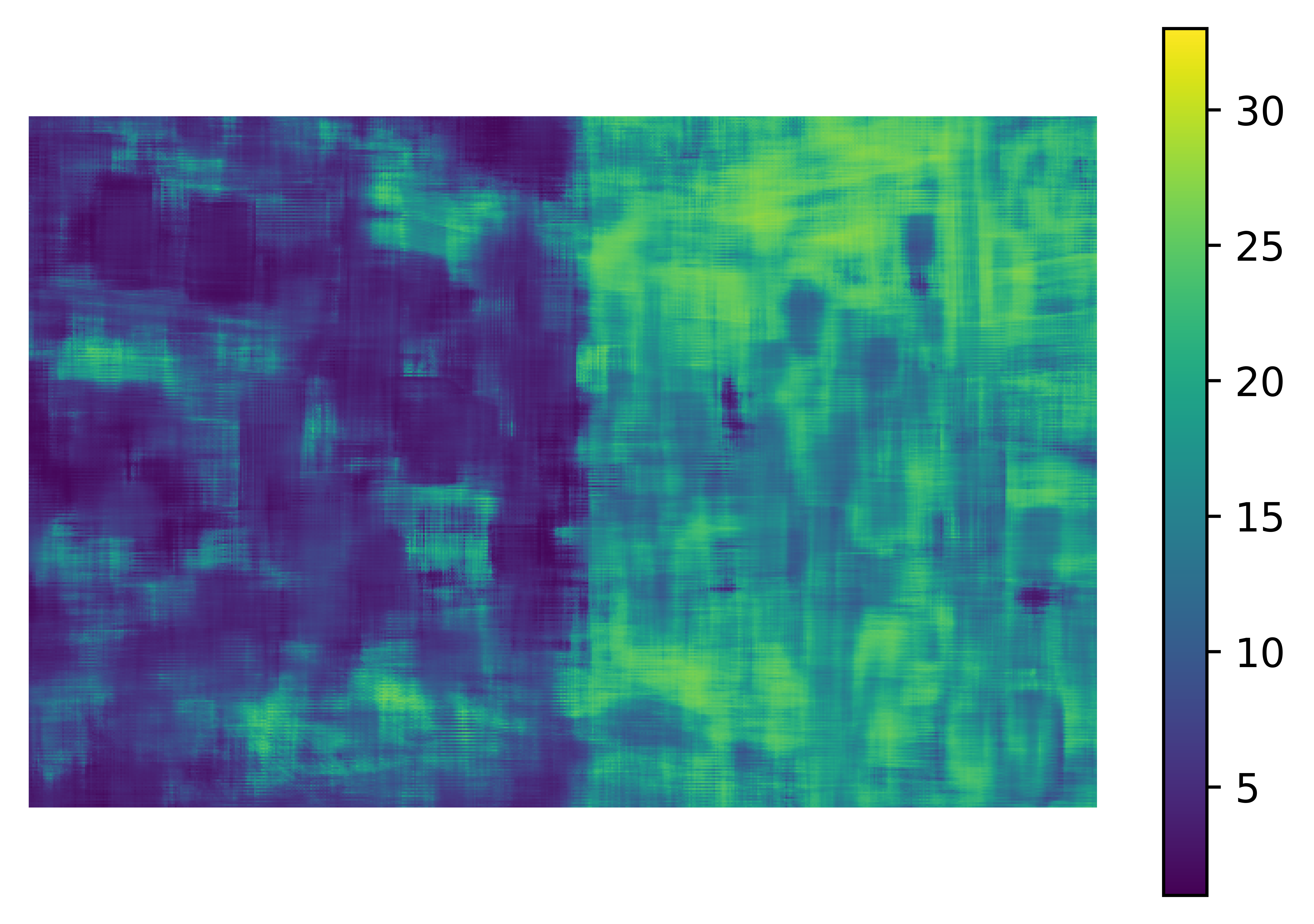}}~~~~~
        {\includegraphics[width=0.3\textwidth]{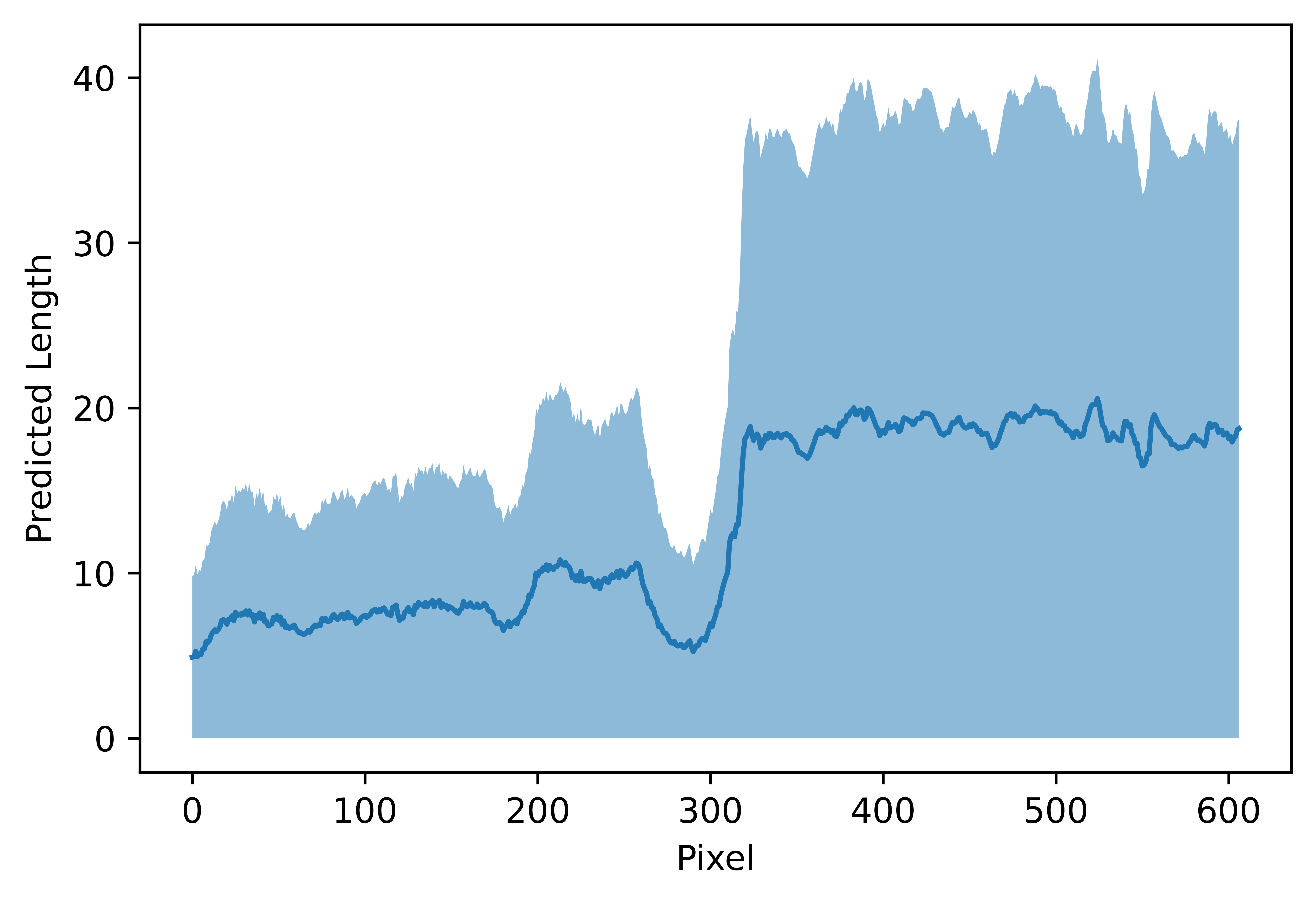}}}\\

        \subcaptionbox{}{{\includegraphics[width=0.3\textwidth]{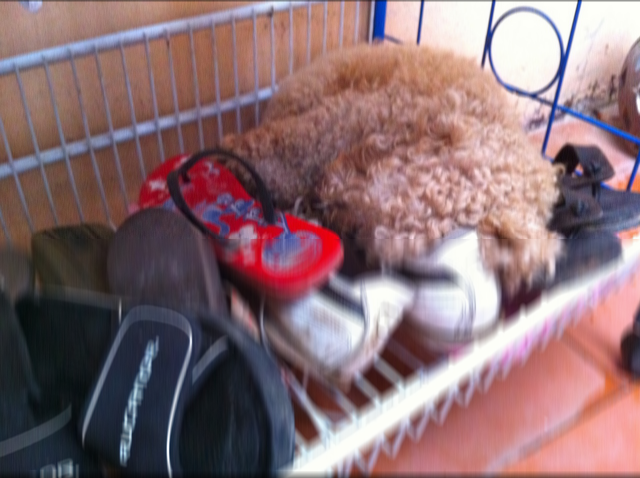}}~~~~~
        {\includegraphics[width=0.3\textwidth]{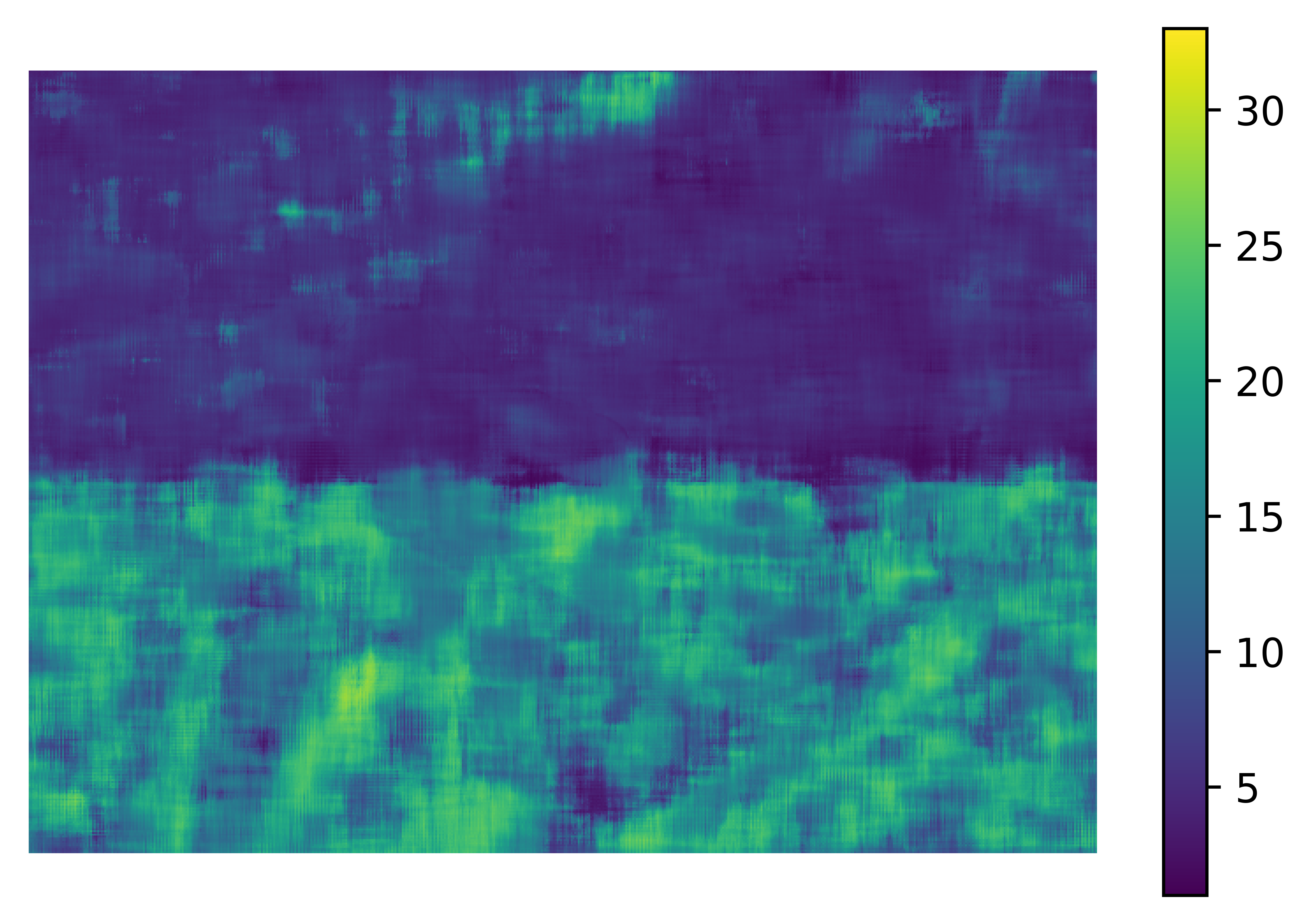}}~~~~~
        {\includegraphics[width=0.3\textwidth]{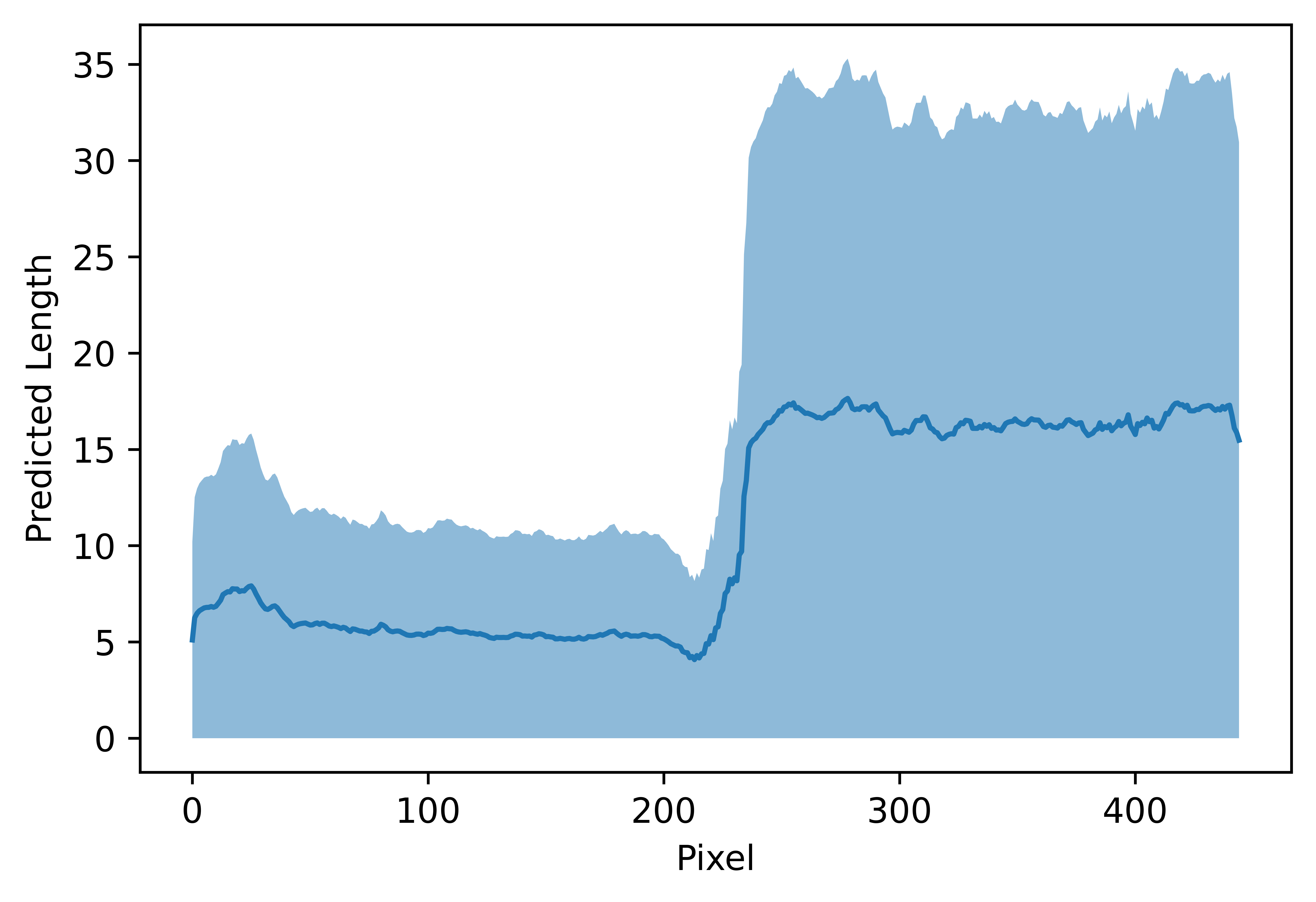}}}\\
        
        \subcaptionbox{}{{\includegraphics[width=0.3\textwidth]{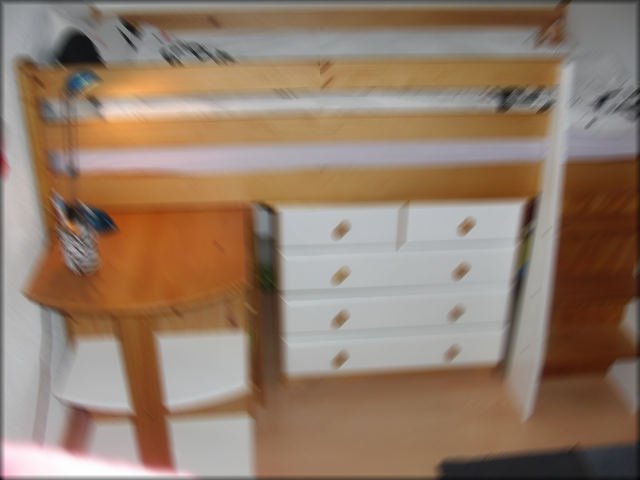}}~~~~~
        {\includegraphics[width=0.3\textwidth]{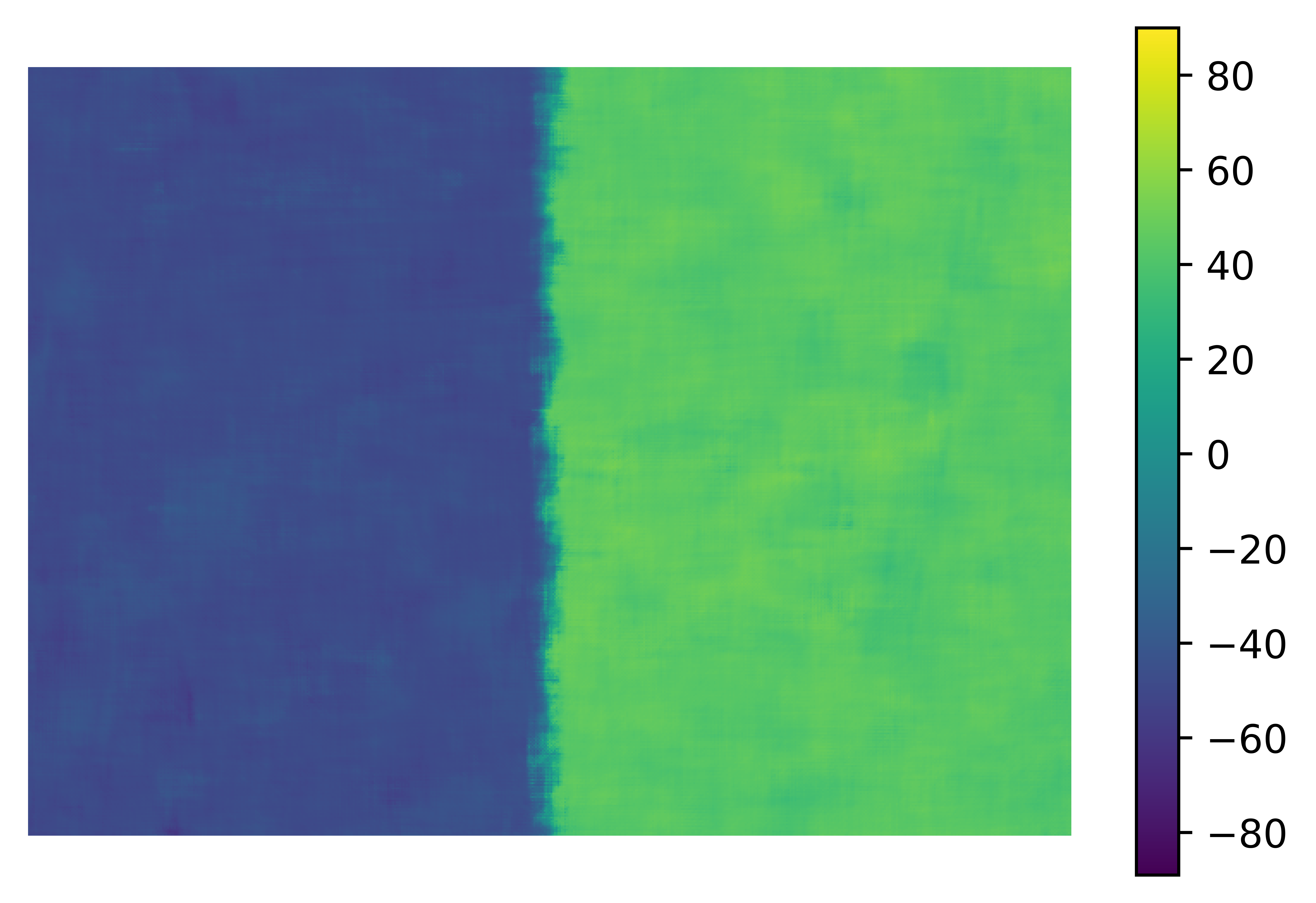}}~~~~~
        {\includegraphics[width=0.3\textwidth]{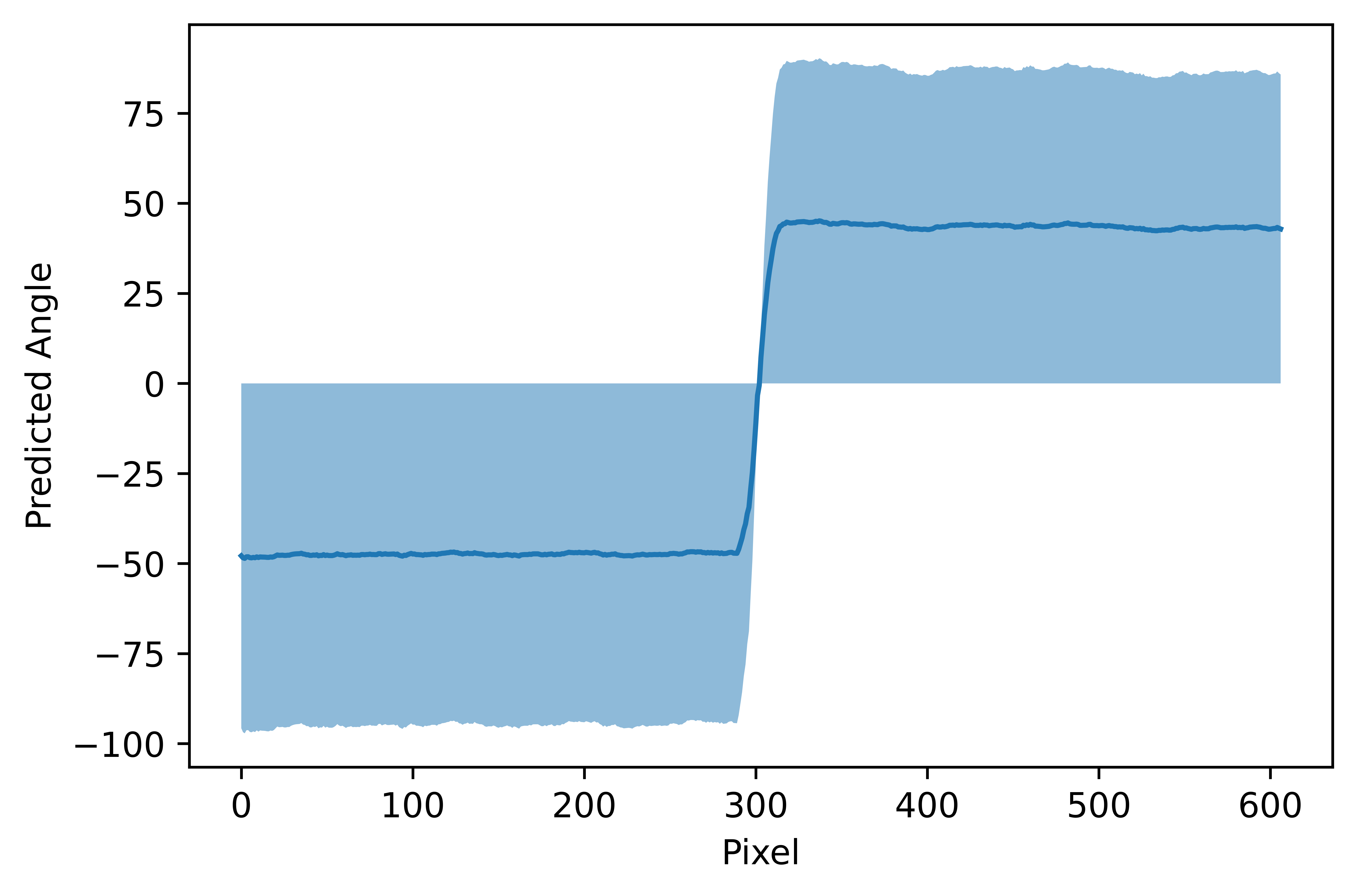}}}\\
        
        \subcaptionbox{}{{\includegraphics[width=0.3\textwidth]{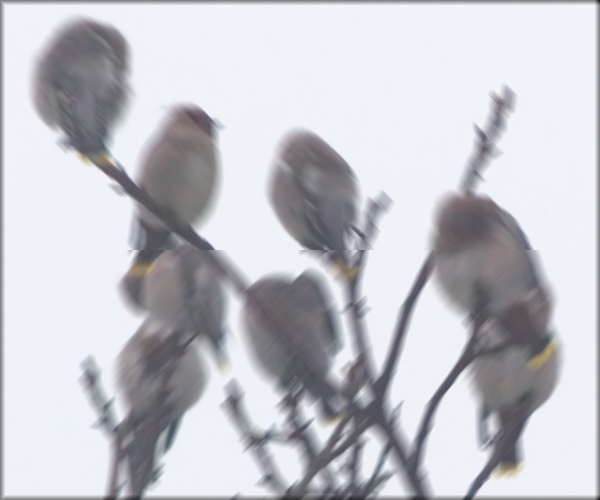}}~~~~~
        {\includegraphics[width=0.3\textwidth]{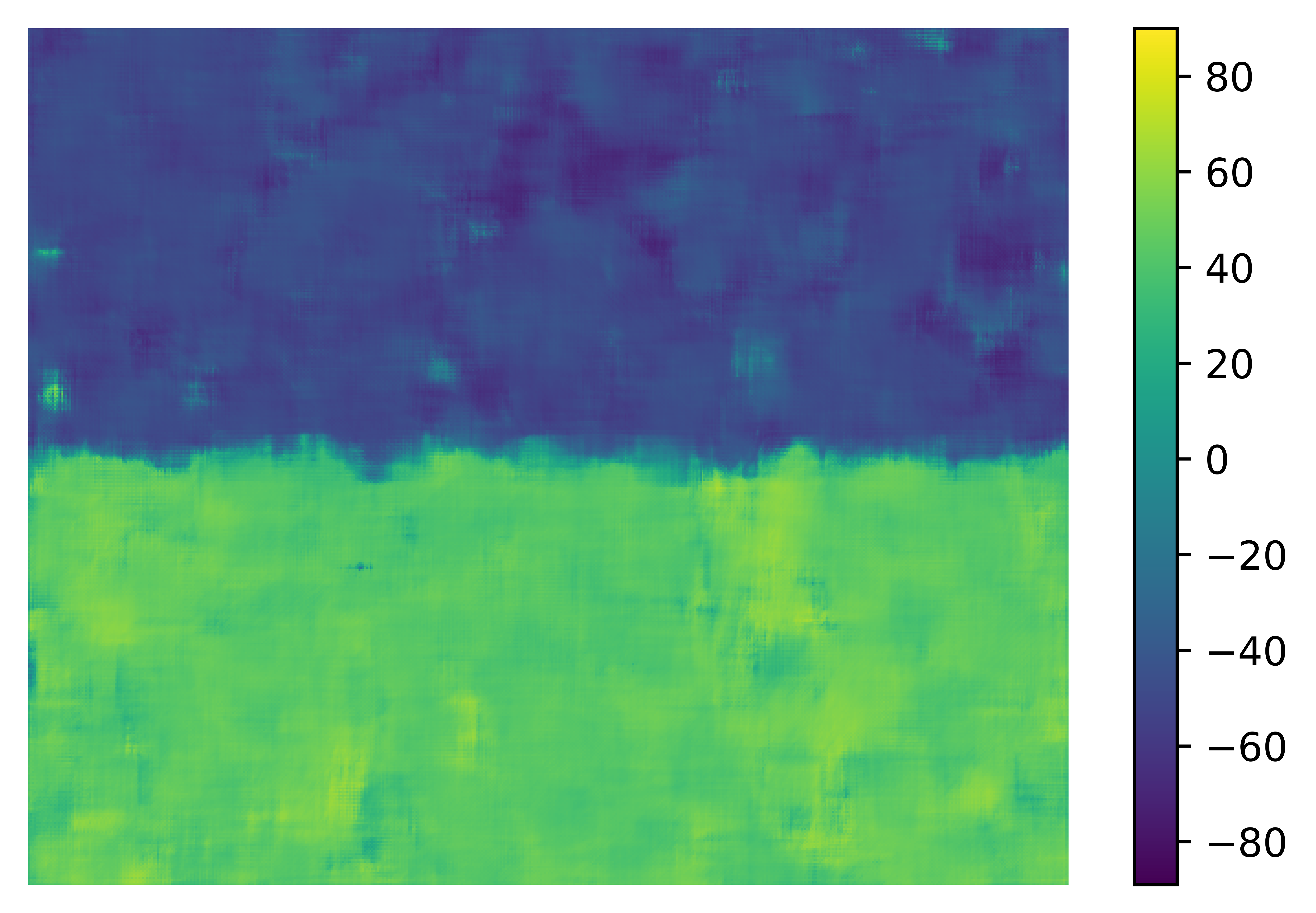}}~~~~~
        {\includegraphics[width=0.3\textwidth]{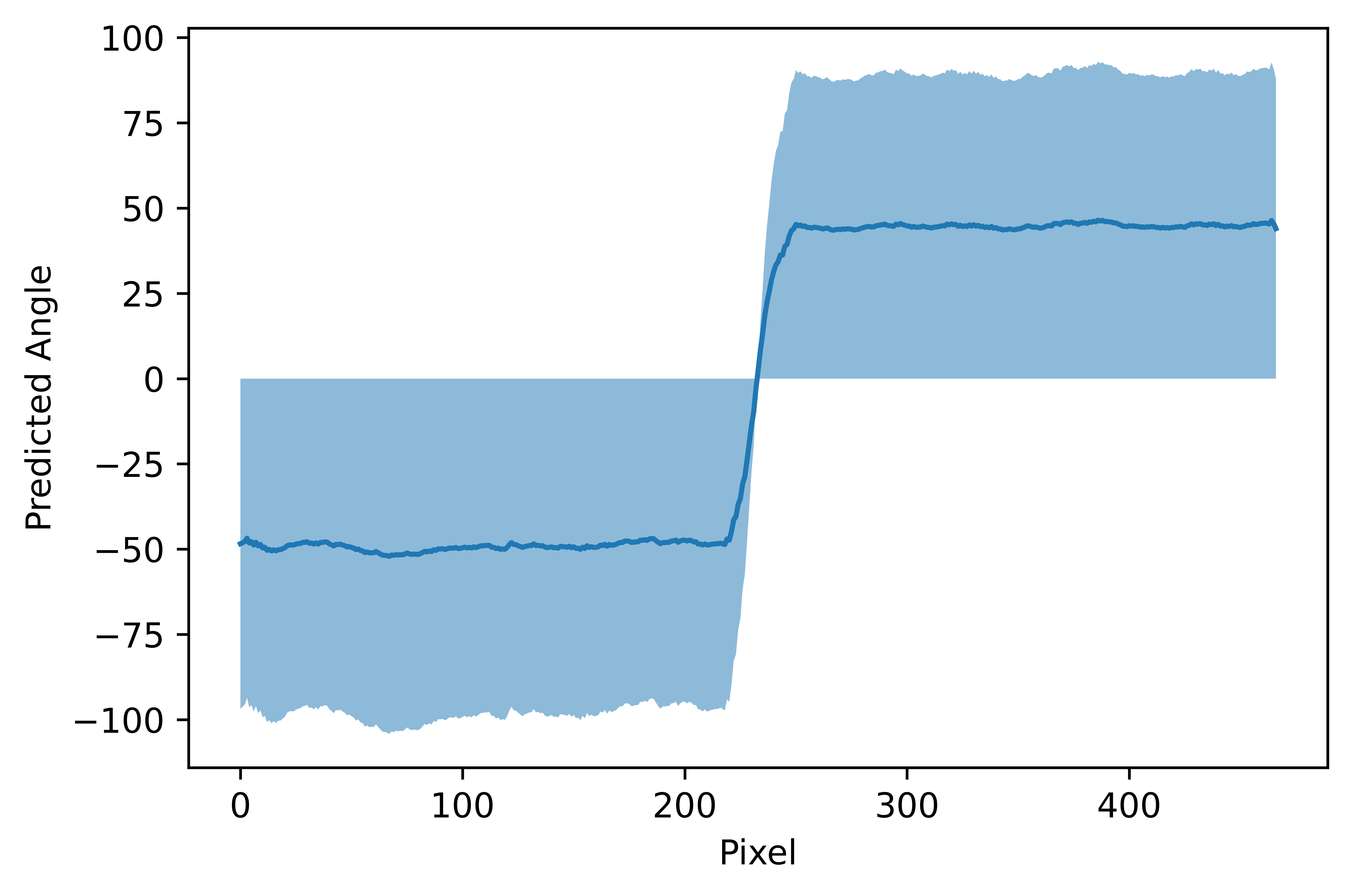}}}\\

        \caption{Predictions of blur characteristics displayed as heatmaps for four example images blurred non-uniformly with (a) \textit{Length Horizontal}, (b) \textit{Length Vertical}, (c) \textit{Angle Horizontal}, and (d) \textit{Angle Vertical}. Left column: the non-uniformly blurred image.  Middle column: the prediction as a heat map.  Note that the heat map is of smaller dimensions than the blurred image since the predictions are patch-based.  Right column: the average (solid line) and standard deviation (shaded area) taken perpendicularly to the blur discontinuity.  Since each image is a different size, the average results in the right column have different scales on the x-axis.} 
        \label{fig:blur_field}
\end{figure}

The intuition that prediction will be between the two represented blur characteristics for a patch overlapping two blur characteristics is represented by the black line in Fig.~\ref{fig:avg_plots}, but the network predictions display some deviation from this intuition, particularly for prediction of length.  The wider variation in the results in Fig.~\ref{fig:avg_plots}(a) are in line with the results summarized by the $R^2$ score, with length prediction resulting in lower $R^2$ scores than angle prediction (see Table~\ref{tab:r2_length}).  The asymmetry in the prediction with respect to the overlap is likely due to the fact that prediction of a larger blur length will require a commensurately larger overlap into that blur characteristic.  From the results in Table~\ref{tab:r2_length}, the network has poorer performance on patches less than $\sim31$.  The results in Fig.~\ref{fig:avg_plots}(a) may be evidence toward the fact that it is prediction of the longer lengths that reduce the $R^2$ score of the network in smaller patch sizes.  We also note some bias in the estimation of length in Fig.~\ref{fig:avg_plots}(a).  We believe this is due to specific image characteristics confounding the length prediction, further discussed below.

In Fig.~\ref{fig:blur_field} we plot a heatmap of predictions for example images blurred with each of the four non-uniform blur patterns discussed in Section~\ref{sec:data}. Consistent with the average results in Fig.~\ref{fig:avg_plots}, we see that the angle predictions [Fig.~\ref{fig:blur_field}(c) and (d)] are more uniform within the two halves of the image as compared to the length predictions [Fig.~\ref{fig:blur_field}(a) and (b)].  The results in Fig.~\ref{fig:blur_field} also give some insight into cases where the network may struggle with correct length prediction.  In particular, many instances of overestimation of length appear to be related to linear structures within the images.  For example, note that the length predictions for the wood floor toward the bottom left of the image in Fig.~\ref{fig:blur_field}(a) are longer than the ground truth value $r=5$.  The structure of the planks in the wood floor, however, is likely confounding the correct prediction of length, providing image structures that look like characteristics of a much longer blur than is applied. 

\section{Conclusion}
\label{sec:conclusion}
This work expands the uniform blur prediction from~\cite{varela2023_uniform} and improves and expands upon the results for a patch-based approach in~\cite{varela2023_patch}.  The training process is modified, alternating different patch sizes per epoch, improving the robustness of the network to varying patch size. Results demonstrate that the network can generalize to a wide range of patch sizes, with better generalization for parameter prediction of length and angle.  Additionally, there is some evidence that a patch size of approximately 30 pixels may be ideal for encompassing important blur cues.

The study of blur characteristic prediction in patches that overlap two regions with different blur indicates that the network predicts blur characteristics in between the overlapped characteristics depending on the level of overlap. Length prediction displayed a wider variance in prediction versus overlap, consistent with the smaller coefficient of determination scores seen for length prediction here and in~\cite{varela2023_uniform,varela2023_patch}.  Observation of predicted blur characteristics in example images indicate that specific linear structures within images are likely the source of errors in length prediction. 

There are several avenues of potential future work.  The removal of the first two convolutional layers produced no significant change in results; future work may study the necessary depth of the network.  Also, it would be interesting to explore the deblurring of the overlapped regions (using approaches similar to those in~\cite{varela2023_uniform}) and analyze artifacts introduced in the deblurring of those overlapped regions. Finally, as our results seem to indicate that a patch size of approximately 30 pixels is sufficient for encompassing important blur cues in images, it could provide important insights to explore other patch sizes.  As other work, e.g.,~\cite{Sun_2015_CVPR, bahat, yan_2016_ieee}, also operate at a similar patch size, specifically training our network toward a smaller patch size, e.g., 16 pixels, may provide additional evidence toward an optimal patch size for non-uniform blur estimation.  Similarly, it may be beneficial to analyze prediction of overlapped regions with smaller and larger patch sizes. 

\section*{Data availability} Data underlying the results presented in this paper for uniformly-blurred images can be generated from the COCO dataset~\cite{lin2014microsoft} via the code available at~\cite{Regression_Blur_github}.  Data underlying the results for non-uniformly blurred images can be generated from the COCO dataset~\cite{lin2014microsoft} via the code available at~\cite{Regression_Patch_Blur_github}.

\section*{Acknowledgments}
The authors gratefully acknowledge Office of Naval Research grant N00014-21-1-2430 which supported this work.

\bibliographystyle{unsrt}
\bibliography{main}

\end{document}